\documentclass[a4paper,11pt]{article}
\usepackage{pos}
\PassOptionsToPackage{colorlinks,linkcolor={blue},citecolor={blue},urlcolor={red}}{hyperref}
\usepackage{hyperref}
\usepackage{natbib}
\usepackage{wrapfig}
\usepackage{float}
\usepackage{color}
\usepackage{graphicx}
\usepackage{gensymb}
\usepackage{amsmath}
\usepackage{enumitem}
\usepackage{url}
\usepackage{multirow,helvet}
\usepackage{todonotes}

\def\mnras{MNRAS}
\def\apj{ApJ}
\def\aj{AJ}
\def\aap{A\&A}

\def\apjl{ApJL}
\def\apjs{ApJS}

\title{Fermi--LAT Gamma-ray Emission Discovered from the Composite Supernova Remnant B0453--685 in the Large Magellanic Cloud}

\author*[a]{Jordan Eagle}

\affiliation[a]{NASA Goddard Space Flight Center,\\
Greenbelt, Maryland, USA}

\emailAdd{jordan.l.eagle@nasa.gov}

\abstract{A second extragalactic pulsar wind nebula (PWN) is discovered in the MeV--GeV band using the {\it Fermi}--LAT. Faint, point-like $\gamma$-ray emission is detected at the location of the composite supernova remnant (SNR) B0453--685 from energies 300\,MeV--2\,TeV. The Fermi--LAT data analysis of the new $\gamma$-ray source is presented together with a detailed multi-wavelength investigation to understand the nature of the observed emission. The observational evidence and physical implications from broadband modeling do not support an SNR $\gamma$-ray origin. Semi-analytic radiative evolutionary models are explored to understand the potential for any pulsar or PWN component responsible for the observed $\gamma$-ray emission. The modeling results favor an evolved PWN ($\tau\sim 14,000$\,years) that has been impacted by the return of the SNR reverse shock with a possible substantial pulsar component below $5\,$GeV. The particle acceleration mechanisms and their efficiency within B0453--685 have important implications for the role PWNe play in generating Cosmic Rays (CRs), but constraints on the synchrotron cut-off are required to accurately characterize the underlying particle properties.}

\FullConference{
Multifrequency Behaviour of High Energy Cosmic Sources XIV (MULTIF2023)\\
12-17 June 2023\\
Palermo, Italy\\}

\begin{document}
\maketitle

\section{Introduction}
\begin{wrapfigure}{r}{0.6\textwidth}
\begin{minipage}{1.0\linewidth}
\vspace{-.65cm}
\hspace{-0.55cm}
\includegraphics[width=1.15\linewidth]{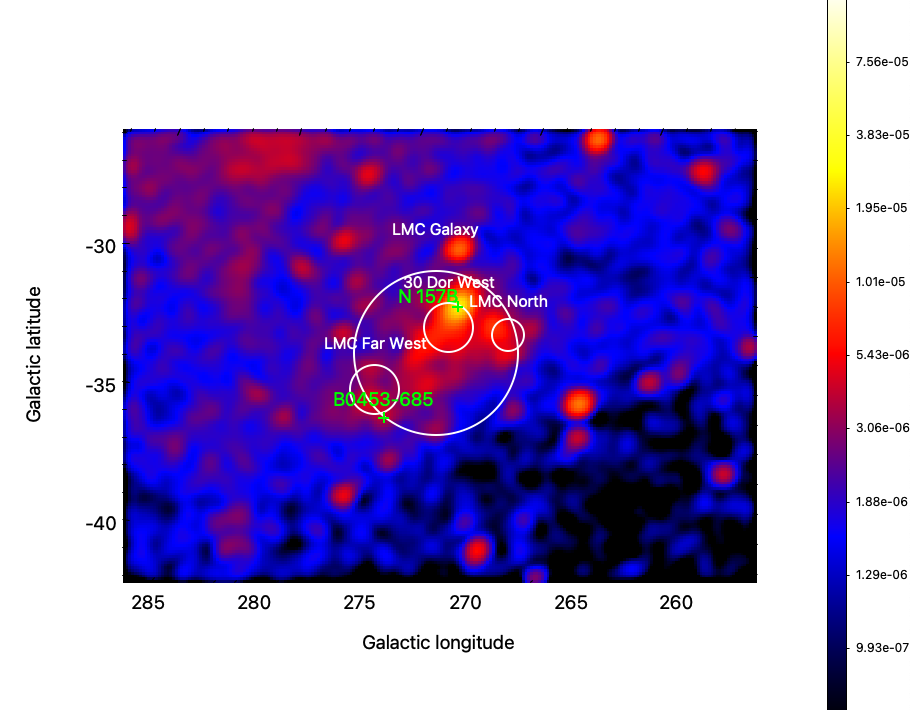}
\end{minipage}
\vspace{-.50cm}
\caption{The Fermi--LAT 12-year counts map (Provided by the Fermi--LAT Collaboration). Sources indicated in green represent Fermi PWNe located within the LMC. See text for details.}\label{fig:fermi_lmc}
\vspace{-.50cm}
\end{wrapfigure}

\begin{wrapfigure}{l}{0.5\textwidth}
\begin{minipage}{1.0\linewidth}
\hspace{-0.35cm}
\includegraphics[width=1.0\linewidth]{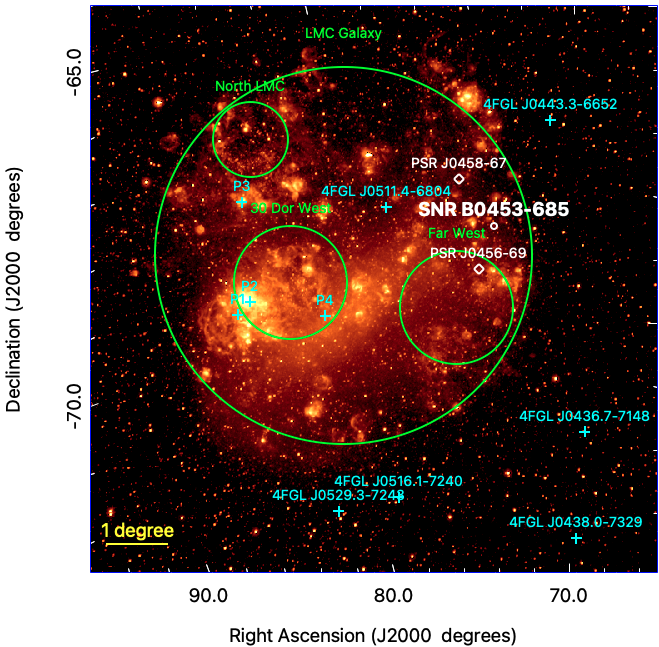}
\end{minipage}
\vspace{-.250cm}
\caption{The LMC in the H$\alpha$ band from the Southern H-Alpha Sky Survey Atlas \citep[SHASSA,][]{halpha2001}. The P1--P4 labels identify the four brightest Fermi point sources in the LMC, following the naming convention used in \citet{lmc2016}. The four extended templates used to describe the diffuse $\gamma$-ray emission from the LMC \citep[components E1--E4 in][]{lmc2016} are indicated with the green circles. The location of SNR~B0453--685 is marked in white with radius $r=0.05\degree$. The two closest known radio pulsars near SNR~B0453--685 are labeled as white diamonds$^a$.}\footnotesize{$^a$ We used the ATNF radio pulsar catalog \url{https://www.atnf.csiro.au/research/pulsar/psrcat/} \citep{manchester2005}.}\label{fig:halpha_lmc}
\vspace{-0.5cm}
\end{wrapfigure}

B0453-685 is located in the Large Magellanic Cloud (LMC), a dwarf satellite galaxy that orbits the Milky Way Galaxy at a distance of 50\,kpc \citep{clementini2003}. Displayed in Figure \ref{fig:fermi_lmc} is the observed Fermi--LAT emission from the LMC for $E > 1\,$GeV (Provided by the Fermi--LAT Collaboration). Only one LMC pulsar wind nebula (PWN), N~157B, is identified as a GeV \citep{4fgl-dr2} and TeV \citep{hess2012} $\gamma$-ray source and it is located on the opposite (Eastern) wall of the LMC with respect to supernova remnant (SNR) B0453--685. N~157B is located in a very crowded area, accompanied by two bright $\gamma$-ray sources nearby, SNR~N132D and PSR~J0540--6919. SNR~B0453--685, however, is conveniently located in a much less crowded region of the LMC, making its faint point-like $\gamma$-ray emission detectable even against the diffuse LMC background, diffuse Galactic foreground, and the isotropic background emissions. Displayed in Figure~\ref{fig:halpha_lmc} is the optical H-alpha emission observed from the LMC by the Southern H-alpha Sky Survey Atlas. Indicated in the figure are the few known sources within the LMC that emit $\gamma$-rays in the Fermi--LAT band, labeled P1--P4 following the convention of \citet{lmc2016}. The cyan sources with 4FGL identifiers are classified as blazars or AGN in the Fermi-LAT catalogs or are unidentified. The green sources mark extended sources that represent the diffuse LMC background observed by the Fermi--LAT. 

\begin{figure*}
\begin{minipage}[b]{.5\textwidth}
\centering
\includegraphics[width=0.85\linewidth]{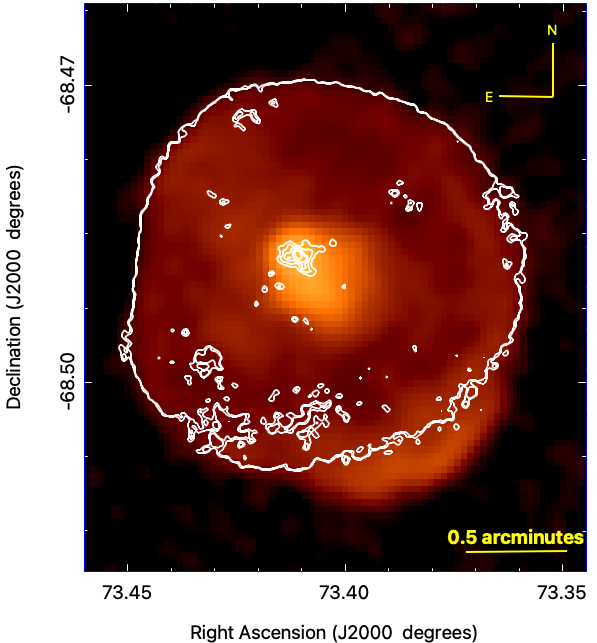}
\end{minipage}
\begin{minipage}[b]{.5\textwidth}
\centering
\includegraphics[width=1.0\linewidth]{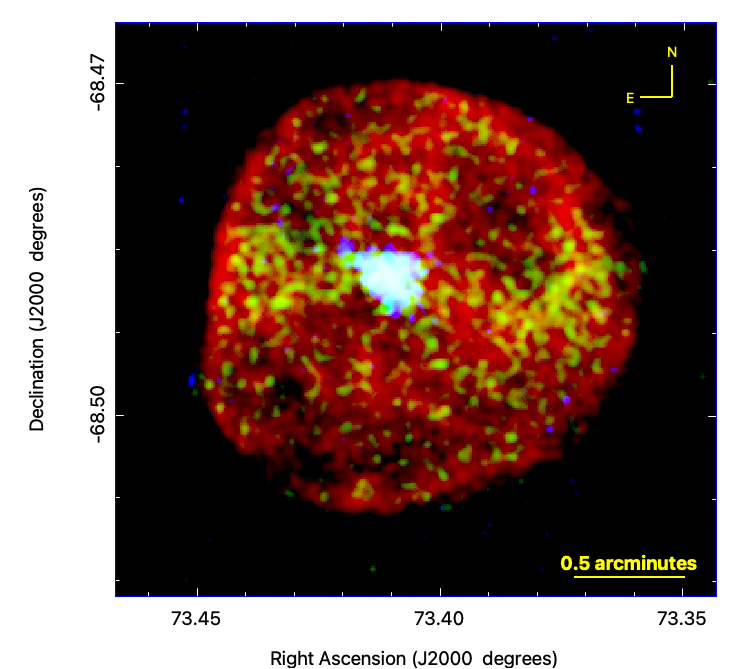}
\end{minipage}
\caption{{\it Left:} The 1.4\,GHz radio emission observed from SNR~B0453--685 \citep{gaensler2003}. The white contours correspond to the central PWN and the outer SNR shell as observed in X-ray (right panel). {\it Right:} Tri-color X-ray flux map of SNR~B0453--685 \citep{gaensler2003}. Red is soft X-ray emission between 0.5--1.2\,keV, green is medium flux between 1.2--2\,keV, and blue is hard flux from 2--8\,keV. Soft and medium X-ray emission outlines and fills the entire SNR while the hard X-ray emission is heavily concentrated towards the center of the SNR where the PWN is located. }\label{fig:radio_xray}
\vspace{-0.5cm}
\end{figure*}

The composite nature of the SNR~B0453-685 was revealed by resolved radio and X-ray observations \citep[][see also Figure~\ref{fig:radio_xray}]{gaensler2003}. A compact centrally polarized central core is observed (the PWN) and a fainter shell ($\sim2\,^\prime$ in diameter) encircling it with center-filled emission visible in both radio and X-ray. The observed morphology indicates B0453-685 is an evolved system where the reverse shock has begun its return to the center of the remnant, reheating the SNR interior as it accelerates toward the PWN. Despite the identification of a PWN and the prediction of a Vela-like pulsar powering it in \citet{gaensler2003}, the central pulsar is undetected. It remains so even after several dedicated pulsar searches targeting the Magellanic Clouds have been performed \citep[e.g.,][]{manchester2006}. 

\section{Fermi--LAT Results}

\begin{figure*}
\begin{minipage}[b]{.5\textwidth}
\centering
\includegraphics[width=1.0\linewidth]{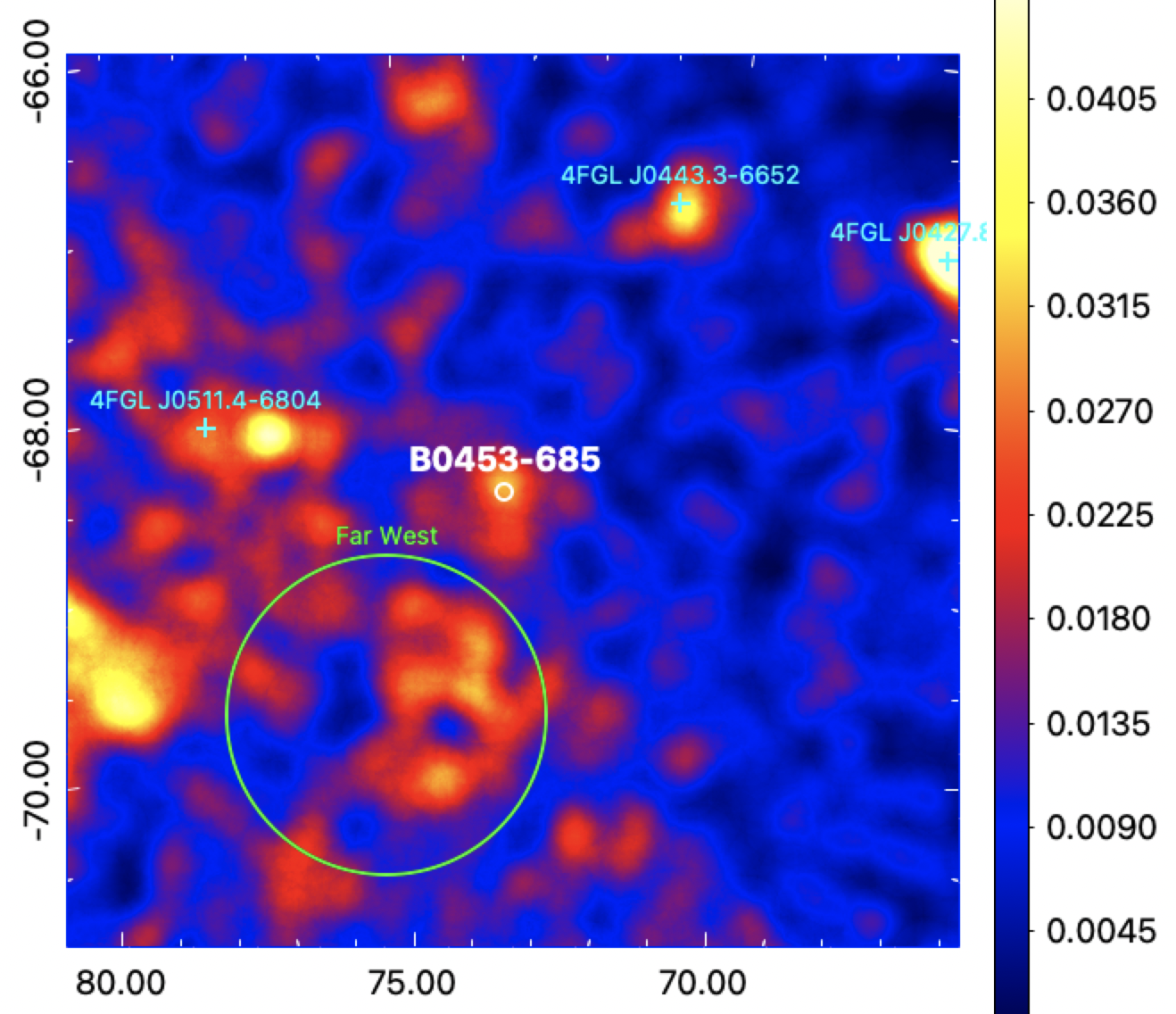}
\end{minipage}
\begin{minipage}[b]{.5\textwidth}
\hspace{-.4cm}
\centering
\includegraphics[width=1.0\linewidth]{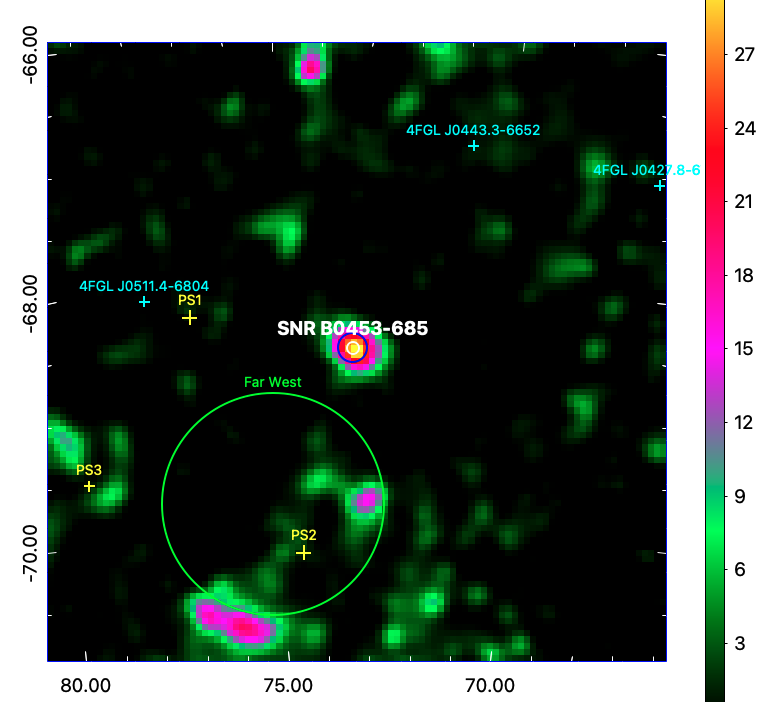}
\end{minipage}
\caption{{\it Left:} Smoothed ($\sigma = 0.1\,\degree$) 5\,$\degree\times 5$\,$\degree$ count map of \texttt{PSF3} events between 1--10\,GeV with the locations of 4FGL sources in the field of view labeled. The pixel size is 0.01 deg pixel$^{-1}$. {\it Right:} 5\,$\degree\times 5$\,$\degree$ TS map between 1--10\,GeV. The maximum TS value at the SNR position is $\sim$ 28 ($\sim 5\sigma$). The 95\% positional uncertainty for the best-fit $\gamma$-ray point source is in blue. In both panels, the location and approximate size of the composite SNR~B0453--685 ($r=0.02\degree$) is marked in white with radius $r=0.05\degree$.}\label{fig:gamma_maps}
\end{figure*}

Using 11.5 years of Fermi-LAT data, faint but significant point-like $\gamma$-ray emission is discovered at the location of B0453-685. This is apparent in both images displayed in Figure~\ref{fig:gamma_maps} where on the left is the 1 to 10 GeV Fermi $\gamma$-ray counts map and on the right is the Test Statistic (TS) map\footnote{The square root of the TS is proportional to the detection significance where a value $TS = 25$ for 1 degree of freedom (DOF) has a $5 \sigma$ significance \citep{mattox1996}.}. The white circle denotes the 2-arcminute diameter and position for the PWN/SNR and the blue circle shows the 95\% positional uncertainty of the $\gamma$-ray source. The point-like nature is consistent with what we would expect if the emission originated from B0453-685, since the Fermi-LAT\footnote{See \url{https://www.slac.stanford.edu/exp/glast/groups/canda/lat_Performance.htm} for a review on the dependence of point-spread-function (PSF) with energy.} can only resolve structure down to $0.1\,\degree$. Most of the $\gamma$-ray signal is observed between energies 1 and 10 GeV and has a soft spectral index $\Gamma_\gamma = 2.3 \pm 0.2$.

\begin{wrapfigure}{r}{0.6\textwidth}
\begin{minipage}{1.0\linewidth}
\vspace{-0.6cm}
\hspace{-0.35cm}
\includegraphics[width=1.1\linewidth]{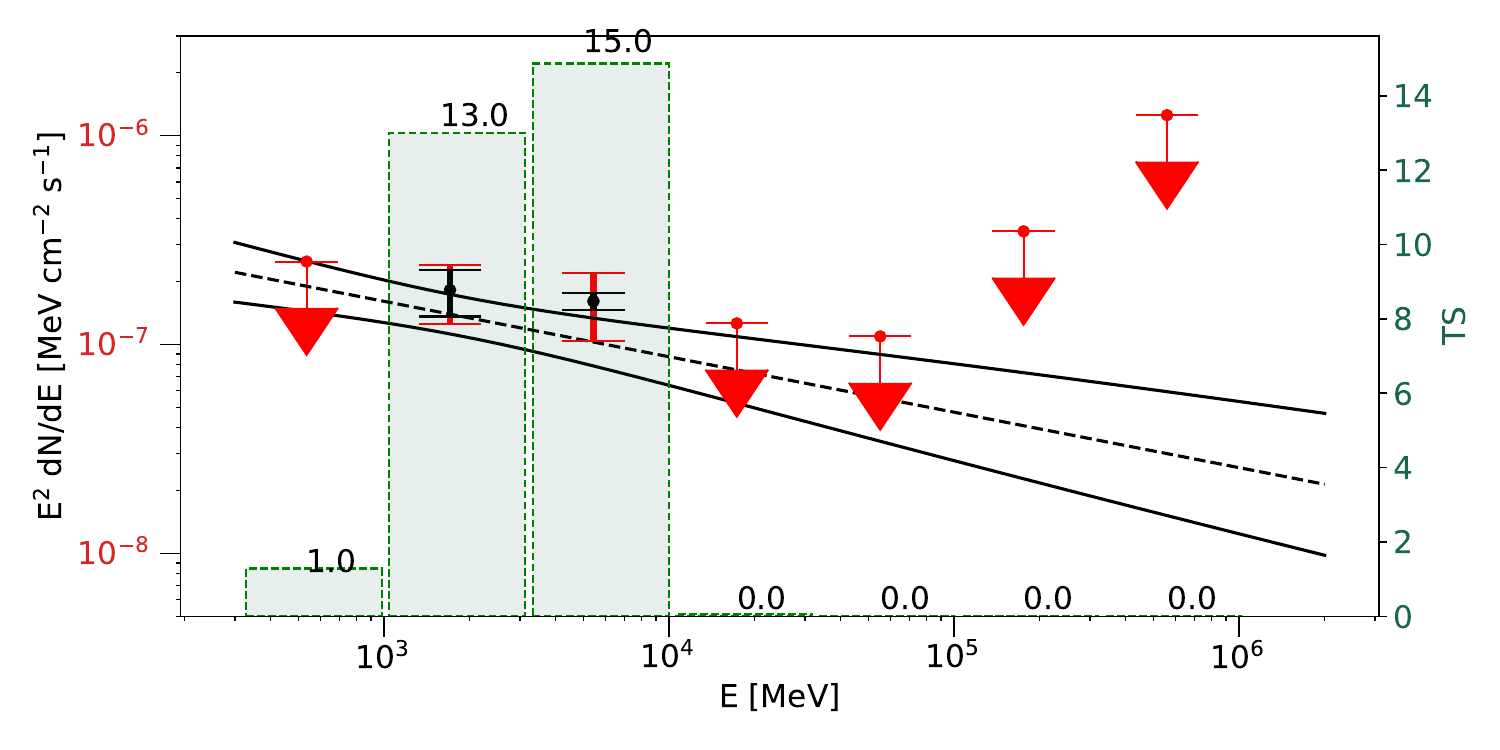}
\end{minipage}
\vspace{-.50cm}
\caption{The best-fit $\gamma$-ray SED for B0453--685 with 1-$\sigma$ statistical uncertainties in red for $TS > 1$ and 95\% confidence level (C.L.) upper limits otherwise. The systematic error from the choice of diffuse LMC model is plotted in black. TS values for each spectral bin are plotted as the green histogram. The data are best characterized as a power-law with $\Gamma = 2.3 \pm 0.2$.}\label{fig:gamma_sed}
\vspace{-0.5cm}
\end{wrapfigure}

Shown in Figure~\ref{fig:gamma_sed} is the best-fit $\gamma$-ray spectral energy distribution (SED) measured for the $\gamma$-ray source. In red are the 1-$\sigma$ statistical errors from the spectral fitting and the black points represent the total systematic error, which accounts for both the uncertainties in the choice of model for the diffuse LMC background and of the LAT instrument performance. As part of a larger systematic study, both the effects from the choice in the diffuse Galactic background models, which mostly impacts the sources along the Galactic plane where the Galactic background dominates, and in the choice for the diffuse LMC background models is considered. To briefly describe the method, the source detection and spectral properties are tested using different background models. This method has been developed and implemented in prior Fermi--LAT catalogs and is probably the most robust way to perform a systematic study with Fermi--LAT data currently \citep[see also e.g.,][]{depalma2013,lmc2016,acero2016}. Unsurprisingly, the effects from the Galactic background model are negligible for the location of the LMC with respect to the Galactic plane. However, we find that the diffuse LMC systematics for B0453-685 behave in a similar way as we see for the diffuse Galactic systematics for sources that lie along the Galactic plane: the systematic uncertainties are comparable to or dominate over the statistical errors for the energy bins with energy lower than 3 GeV and become negligible for the bins above this energy. 

\section{$\gamma$-ray Origin through Broadband Modeling}
Both the PWN and SNR shell are plausible counterparts based on the positional coincidence, as well as the central pulsar, even though undetected, may still be a plausible origin. Given the intrinsic faintness of the $\gamma$-ray emission, a blind pulsation search is not feasible. Instead, we focus our efforts on combining the Fermi-LAT data to available multi-wavelength data for this system and perform broadband modeling. \citet{haberl2012} reported both the PWN or ``core'' and the SNR shell radio spectral measurements for various radio bands, and archival {\it Chandra} X-ray observations were re-analyzed to extract the PWN spectrum, which is very hard with a measured photon index of 1.7, see Figure~\ref{fig:xray_sed}. No nonthermal X-ray emission is attributed to the SNR, only thermal X-rays.  We extract an upper limit for the broadband modeling which is performed using the NAIMA Python package \citep{naima}. We test a power-law with an exponential cut-off particle distribution,
\begin{equation}
f(E) = A \bigg(\frac{E}{E_0}\bigg)^{-\Gamma} \exp\big({-{\frac{E}{E_{c}}}}\big) \\
\end{equation}
allowing the normalization, first index, cut-off energy, and the magnetic field for the synchrotron component to vary. 

\subsection{SNR Origin}

Table~\ref{tab:obs_constraints} provides the observational constraints used to motivate and interpret the models alongside the predicted properties from each model. The total particle energies $W_e$, $W_p$, and the magnetic field strength $B$ are results of the best-fit for two cases for the SNR: hadronic-dominant or leptonic-dominant. The best-fit models are displayed in the lower panels of Figure~\ref{fig:mybroadbandpwnmodel}.  The average value of the coherent magnetic field which is expected to be the dominant component at this scale is estimated to be 1.0\,$\mu$G \citep{gaensler2005}. The average particle density for the interstellar medium (ISM) of the LMC is measured as 2.0\,cm$^{-3}$ \citep{kim2003}. If the SNR were efficiently accelerating particles, it is likely that shock compression is amplifying both of these quantities by at least a factor of 4 \citep[see e.g.,][and references therein]{castro_2013}. 

\begin{wrapfigure}{l}{0.53\textwidth}
\begin{minipage}{1.0\linewidth}
\vspace{-0.6cm}
\hspace{-0.35cm}
\includegraphics[width=1.05\linewidth]{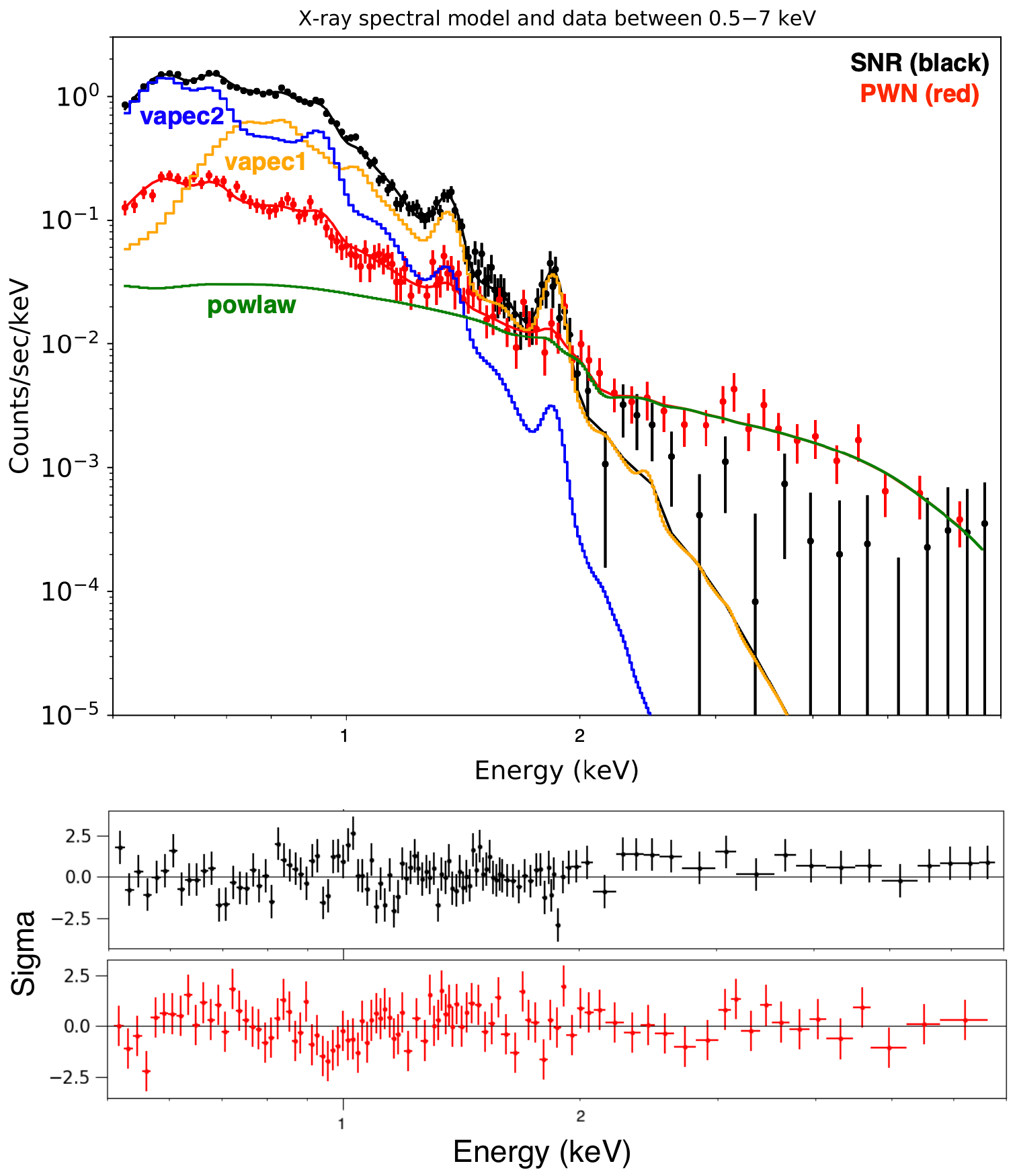}
\end{minipage}
\vspace{-.250cm}
\caption{{\it Chandra} X-ray spectral data and models. The PWN is best-fit as a simple power law, $\Gamma_\gamma \approx 1.7$. Adapted from \citet{eagle2023}.}\label{fig:xray_sed}
\vspace{-1.25cm}
\end{wrapfigure}

 An amplified magnetic field would increase the synchrotron emission at the SNR shock front, which we do not detect from B0453-685 (Figure~\ref{fig:xray_sed}). Secondly, the post-shock density estimated from the LMC average density, assuming a compression ratio of 4, gives a very large value $n_H = 8.0$\,cm$^{-3}$, in contrast to the one inferred from X-ray observations for the ambient medium around B0453-685, $n_H = 1.6$\,cm$^{-3}$. Observations in both X-ray and optical (Figure~\ref{fig:halpha_lmc}) indicate a lower density in the region of B0453--685 than the rest of the LMC. This is consistent with the observed faint SNR shell in both X-ray and radio (Figure~\ref{fig:radio_xray}), implying that the shell is not energetically interacting with denser material. Finally, the total particle energy in both SNR scenarios requires a significant fraction of the supernova explosion energy be carried away by the particles. X-ray observations infer a supernova explosion energy on the order of $E_{SN} \sim 10^{50}\,$erg \citep{gaensler2003}. In summary, the results of the radiative modeling, the age of the SNR, and the lack of non-thermal X-rays from the shell do not support an SNR origin. 

\subsection{PWN Origin}

\begin{figure*}[htbp]
\begin{minipage}[b]{1.0\textwidth}
\centering
\includegraphics[width=0.75\linewidth]{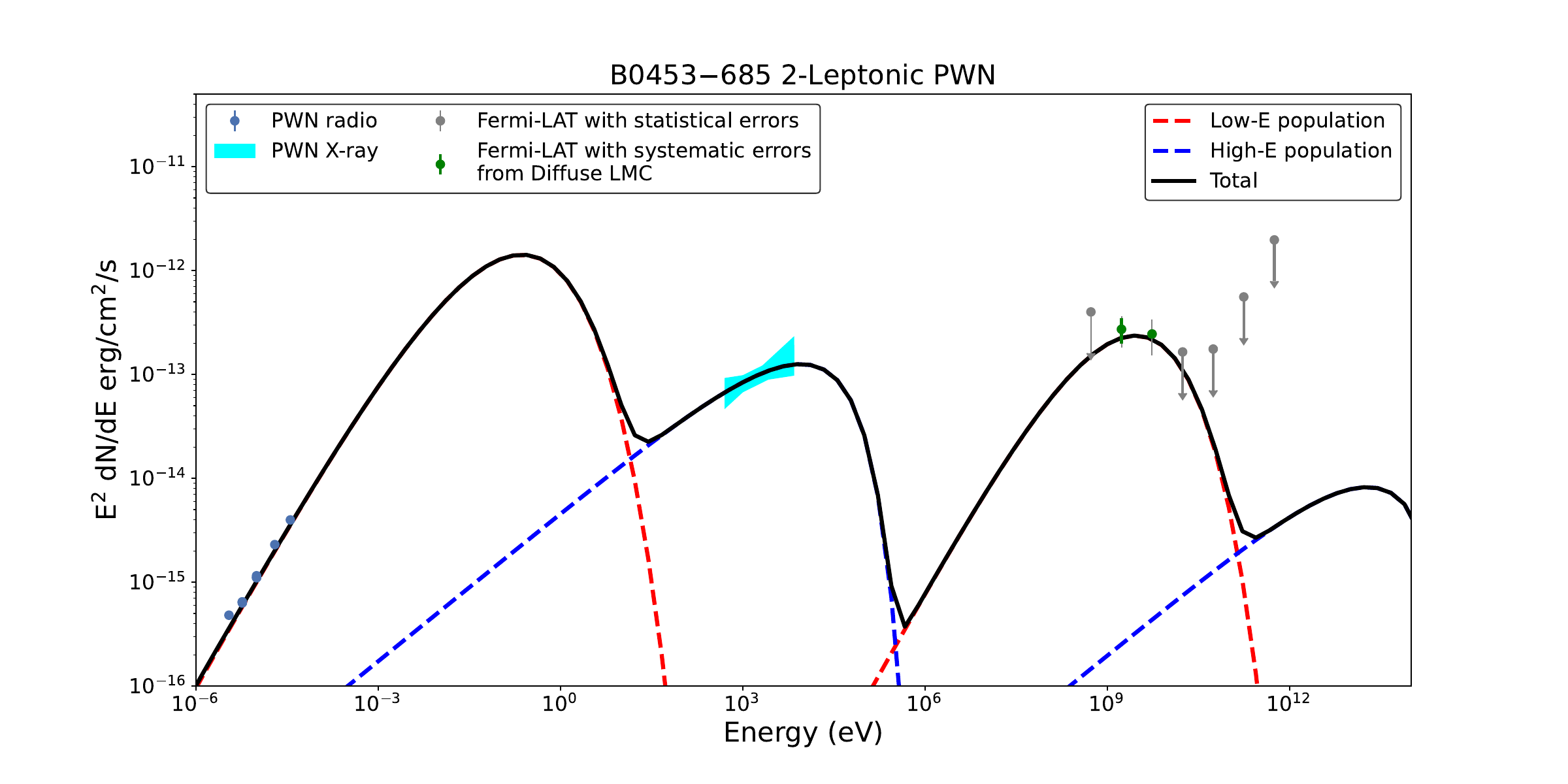}
\vspace{-0.35cm}
\end{minipage}
\begin{minipage}[b]{1.0\textwidth}
\centering
\includegraphics[width=1.0\linewidth]{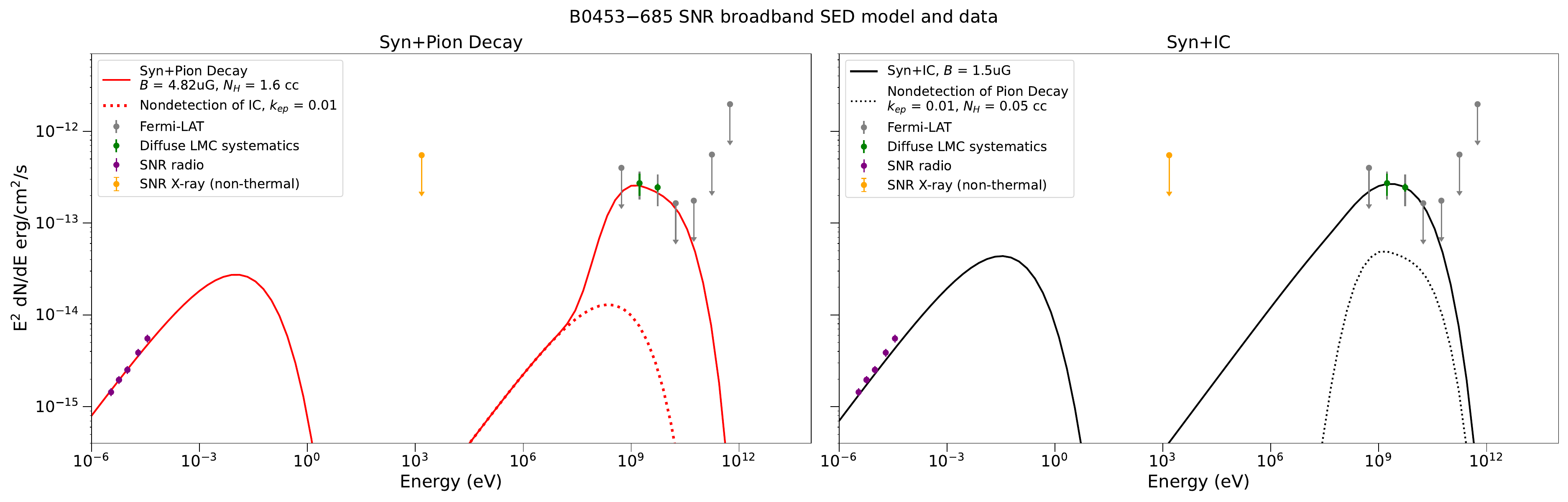}
\end{minipage}
\vspace{-0.5cm}
\caption{The best-fit broadband models for the three scenarios investigated to understand $\gamma$-ray origin. {\it Top:} Two leptonic populations are required to explain the broadband PWN emission. {\it Bottom Left:} a single leptonic population describing SNR synchrotron emission combined with a single hadronic population describing the $\gamma$-ray emission via pion decay from the SNR. {\it Bottom Right:} The case where the leptonic population dominates over the hadronic population in the SNR. 
Radio data of PWN (blue) and SNR (purple) are from \citet{haberl2012}. X-ray data of PWN is in cyan and for the SNR it is yellow, and the $\gamma$-ray data is shown in grey/green. Adapted from \citet{eagle2023}.}\label{fig:mybroadbandpwnmodel}
\end{figure*}

For the PWN case, we invoke the same particle distribution shape but incorporate two leptonic populations in order to explain the broadband emission. This is consistent with several evolved PWNe which also require the presence of more than one leptonic population \citep[e.g.,][]{,eagle2022}. Based on simple radiative modeling results presented in Table~\ref{tab:obs_constraints} and the top panel of Figure~\ref{fig:mybroadbandpwnmodel}, a PWN $\gamma$-ray origin seems the most likely. 

\section{PWN Origin through Radiative Evolutionary Modeling}
We can attempt to predict other basic energetics of the system incorporating the energy loss as the PWN evolves. Many radiative evolution codes have been developed to predict several physical quantities of the system and the one implemented here is described in \citet{gelfand_2009}. Here, we consider a time-dependent particle distribution which takes the shape of a broken power-law,
\begin{equation}
\frac{d\dot{N}_{e^\pm}(E)}{dE} =
\begin{cases}
 \dot{N}_{break} \big(\frac{E}{E_{break}}\big)^{-p_1} & E_{min} < E < E_{break} \\
 \dot{N}_{break} \big(\frac{E}{E_{break}}\big)^{-p_2} & E_{break} < E < E_{max} \\
\end{cases}
\end{equation}
where $\dot{N}_{e^\pm}$ is the rate that electrons and positrons are injected into the PWN, and $\dot{N}_{break}$ is calculated using
\begin{equation}
  \eta_P\dot{E} = \int_{E_{min}}^{E_{max}} E \frac{d\dot{N}(E)}{dE} dE
\end{equation}
The particle injection rate depends on the underlying particle properties, the progenitor and ambient medium properties, as well as the pulsar characteristics, namely the spin-down power $\dot{E}(t)$ and the age $t_{age}$,
\begin{equation}
  t_{age} = \frac{2t_{ch}}{p-1} - \tau_{sd}
\end{equation}
and the spin-down luminosity $\dot{E}$ is defined as
\begin{equation}
  \dot{E}(t) = \dot{E_0}\big(1 + \frac{t}{\tau_{sd}}\big)^{-\frac{p+1}{p-1}}
\end{equation}
and are chosen for a braking index $p$, initial spin-down luminosity $\dot{E_0}$, and spin-down timescale $\tau_{sd}$ to best reproduce the pulsar's likely characteristic age and current spin-down luminosity. A fraction of pulsar rotational energy is transferred to the magnetosphere, the rest $(1-\eta_\gamma)$ is injected into the PWN in the form of a magnetized, highly relativistic outflow, i.e., the pulsar wind. The pulsar wind enters the PWN at the termination shock, where the rate of magnetic energy $\dot{E}_B$ and particle energy $\dot{E}_P$ injected into the PWN is expressed as:
\begin{eqnarray}\label{eqn:edotb}
\dot{E}_B(t) & \equiv & \eta_{\rm B}\dot{E}(t) \\
\dot{E}_P(t) & \equiv & \eta_{\rm P}\dot{E}(t)
\end{eqnarray}
where $\eta_B$ is the magnetization of the wind and defined to be the fraction of the pulsar's spin-down luminosity injected into the PWN as magnetic fields and $\eta_P$ is the fraction of spin-down luminosity injected into the PWN as particles. The radiative evolution code predicts the properties of the supernova progenitor, the ambient medium, and properties of the PWN and PSR once the model reasonably reproduces the observational constraints (e.g., PWN/SNR size and broadband data).

\begin{figure*}
\begin{minipage}[b]{.5\textwidth}
\centering
\includegraphics[width=1.0\textwidth]{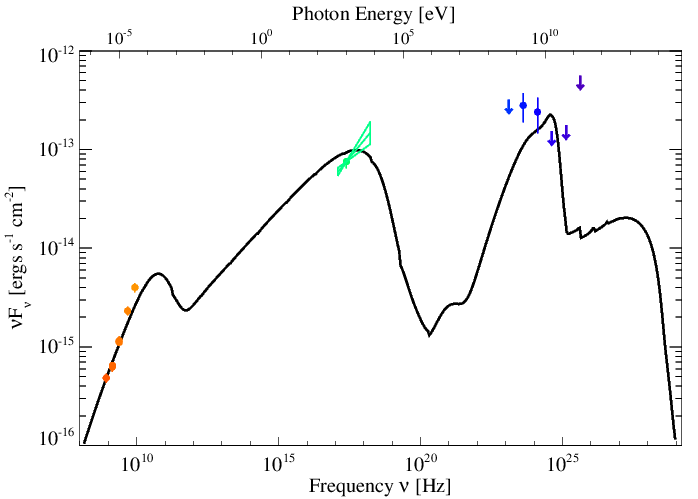}
\end{minipage}
\begin{minipage}[b]{.5\textwidth}
\centering
\includegraphics[width=1.0\textwidth]{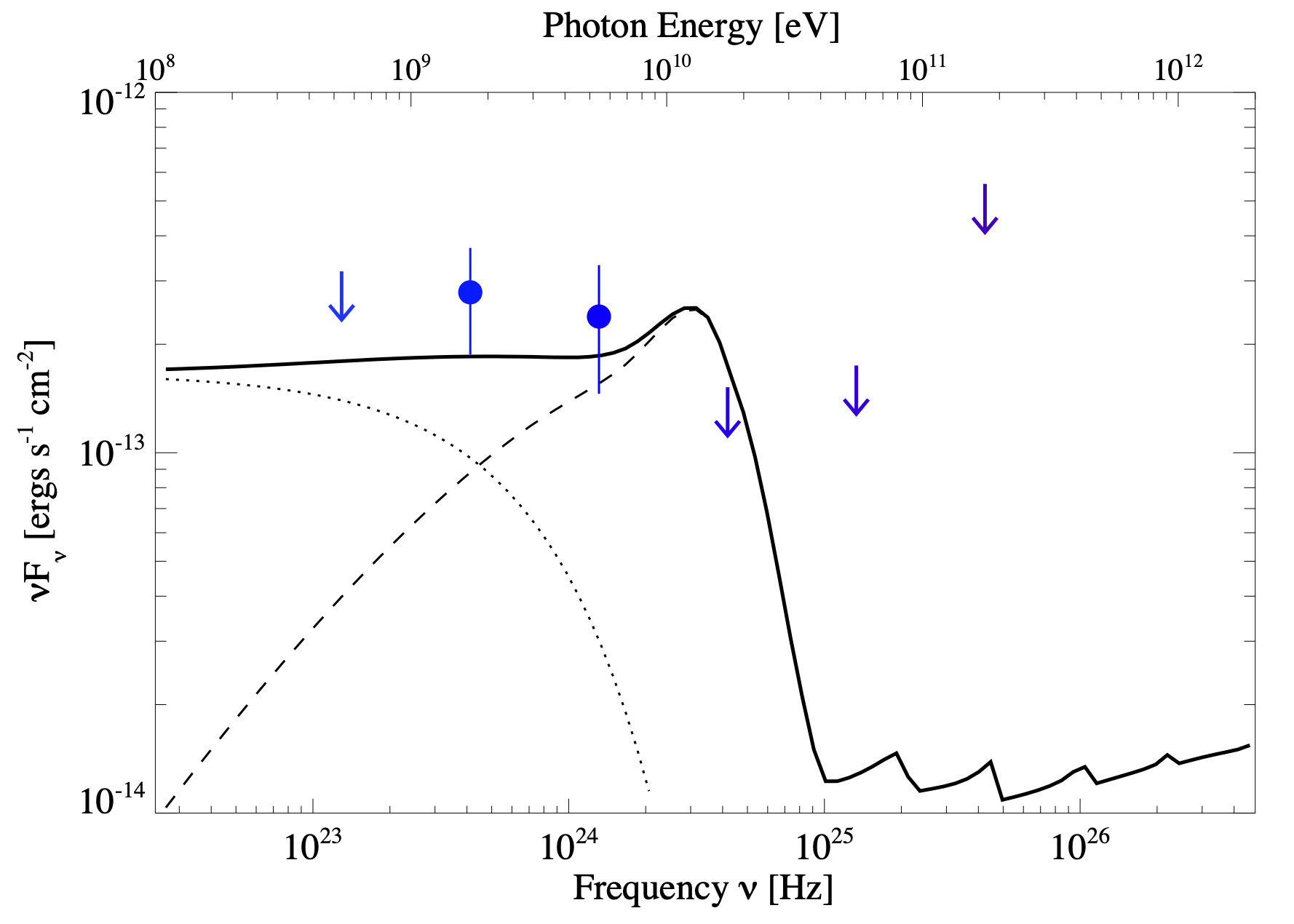}
\end{minipage}
\caption{{\it Left:} The best-fit SED assuming all Fermi--LAT emission is non-magnetospheric in origin (i.e., PWN only). {\it Right:} The $\gamma$-ray spectral evolutionary model assuming magnetospheric contribution to the Fermi--LAT emission. The dotted line indicates the pulsar contribution and the dashed line indicates the PWN contribution. The colored points represent the values of observed data that the model used as comparison points for fitting and are the same values as those in the top panel of Figure~\ref{fig:mybroadbandpwnmodel}. In both panels, the discontinuous spectral features beyond $\nu \sim 10^{25}\,$Hz are numerical artifacts and can be ignored.}\label{fig:compare-gelfand-seds}
\end{figure*}

The final results of the best-fit evolutionary broadband model for the PWN is displayed in Figure~\ref{fig:compare-gelfand-seds}, left panel. We see that the predicted age $\tau \sim 14.3$\,kyr, ambient particle density $n_0 \sim 1.0$\,cm$^{-3}$, and supernova explosion energy $E_{SN} \sim 5\times10^{50}\,$erg, listed in Table~\ref{tab:inputoutputgelfand}, are in reasonable agreement with those inferred by the X-ray observations (Table~\ref{tab:obs_constraints}). However, the $\gamma$-ray data is not sufficiently modeled in the left panel of Figure~\ref{fig:compare-gelfand-seds}. Adding a second spectral component that is very typical for Fermi-detected pulsars \citep[e.g.,][]{2pc}, a power-law photon spectrum with an exponential cut-off somewhere between 1 and 10\,GeV, can better characterize the observed $\gamma$-ray spectrum. This second component, shown as a dotted line in the right panel of Figure~\ref{fig:compare-gelfand-seds}, together with the PWN contribution (dashed) provides the best fit to the Fermi--LAT emission. In this scenario, the PWN only becomes dominant above a few GeV. The full set of predicted properties for the injection spectrum, pulsar, and PWN are listed in Table~\ref{tab:inputoutputgelfand}.

\begin{table*}[tbh!]
\scalebox{0.85}{
\hspace{-0.5cm}
\begin{tabular}{| c || c c || c c c |}
\hline
\ & \multicolumn{2}{|c||}{Observed} & \multicolumn{3}{|c|}{Predicted} \\
\hline
\hline
\ Property & LMC & B0453-685 & Two-Leptonic PWN & Hadronic-dominant SNR & Leptonic-dominant SNR \\
\hline 
\hline 
\  B ($\mu$\,G) & 1.0 & -- & 8.2 & 4.8 & 1.5 \\
\hline 
\ $n_0$ (cm$^{-2}$) & 2.0 & 0.4 & -- & 0.4 & 0.01 \\
\hline 
\ $n_H$ (cm$^{-2}$) & 8.0 & 1.6 & -- & 1.6 & 0.05 \\
\hline 
\ $\tau$ (kyr) & -- & 13.0 & -- & -- & -- \\
\hline 
\ $v_s$ (km s$^{-1}$) & -- & $\lesssim 500$ & & -- -- & -- \\
\hline 
\ $W_e$ or $W_p$ ($10^{51}$ erg) & -- & 0.7$^a$ & 0.03 & 4.0 & 0.3 \\
\hline 
\end{tabular}}
\caption{Observational constraints derived for the LMC from various studies \citep{kim2003,gaensler2005} and B0453--685 properties inferred from X-ray observations \citep{gaensler2003} compared to the predicted properties from the two-leptonic PWN model and each presented SNR model. $v_s$ is the shock velocity of the SNR shell, estimated assuming a Sedov-Taylor solution setting $\tau = 13$\,kyr. $n_0$ is the pre-shock proton density and $n_H$ is the post-shock proton density. \footnotesize{$^a$ is the SN explosion energy $E_{SN}$.}}
\label{tab:obs_constraints}
\end{table*}

\section{Particle Acceleration Prospects for PWN~B0453--685}
The physical implications for the underlying particle properties is interesting to consider in the context of PWNe being efficient particle accelerators. As a general comparison, the Crab nebula also requires two electron populations to most accurately reproduce the broadband data, with a high-energy particle index of 2.2 and a low-energy particle index of 1.6 \citep{lyutikov2019}, which is similar to our values of 2.4 and 1.3 shown in Table~\ref{tab:naima_ev} where we compare the major differences between the NAIMA best-fit representation and the one derived from the evolutionary code. \citet{lyutikov2019} derived a theoretical interpretation of the underlying particle populations for the Crab: the high-energy particles, which dominate in the MeV-GeV $\gamma$-ray band, have a soft index and are undergoing diffusive shock acceleration mainly within the equatorial region where the termination shock is located. The low-energy particles dominate the radio emission, have a harder index, and are not likely to undergo diffusive shock acceleration, but may be efficiently accelerated within the polar regions of the nebula by magnetic turbulence such as magnetic reconnection and/or Weibel instabilities \citep[e.g.,][]{sironi_2011}. See Figures~4, 5 and 6 of \citet{lyutikov2019} for additional information.

\begin{table*}
\scalebox{0.9}{
\hspace{-0.25cm}
\begin{tabular}{|c c c c|}
\hline
\/Shorthand & Parameter & PWN$+$PSR Best-Fit & Units \\
\hline
\hline
\ \texttt{loglh} & Log Likelihood of Spectral Energy Distribution  & --17.6 & -- \\
\hline
\ \texttt{esn} & Initial Kinetic Energy of Supernova Ejecta  & 5.21 & $10^{50}$ ergs \\
\hline
\ \texttt{mej} & Mass of Supernova Ejecta & 2.42 & Solar Masses \\
\hline
\ \texttt{nism} & Number Density of Surrounding ISM & 1.00 & cm$^{-3}$ \\
\hline
\ \texttt{brakind} & Pulsar Braking Index & 2.83 & - \\
\hline
\ \texttt{tau} & Pulsar Spin-down Timescale & 166 & years \\
\hline
\ \texttt{age} & Age of System & 14300 & years \\
\hline
\ \texttt{e0} & Initial Spin-down Luminosity of Pulsar & 6.79 & $10^{39}$ ergs s$^{-1}$ \\
\hline
\ \texttt{etag} & Fraction of Spin-down Luminosity lost as Radiation & 0.246 & - \\
\hline
\ \texttt{etab} & Magnetization of the Pulsar Wind & 0.0007 & - \\
\hline
\ \texttt{emin} & Minimum Particle Energy in Pulsar Wind & 2.26 & GeV \\
\hline
\ \texttt{emax} & Maximum Particle Energy in Pulsar Wind & 0.73 & PeV \\
\hline
\ \texttt{ebreak} & Break Energy in Pulsar Wind & 72 & GeV \\
\hline
\ \texttt{p1} & Injection Index below the Break & 1.34 & - \\
\ & (${dN}/{dE} \sim E^{-p1}$) & & \\
\hline
\ \texttt{p2} & Injection Index below the Break & 2.36 & - \\
\ & (${dN}/{dE} \sim E^{-p2}$) & & \\
\hline
\ \texttt{ictemp} & Temperature of each Background Photon Field & 1.13 & $10^{6}$ K \\
\hline
\ \texttt{icnorm} & Log Normalization of each Background Photon Field & -18.0 & - \\
\hline
\ \texttt{gpsr} & Photon Index of the $\gamma$-rays Produced Directly by the Pulsar & 2.00 & -- \\
\hline
\ \texttt{ecut} & Cutoff Energy from the Power Law of Pulsar Contribution & 3.21 & GeV \\
\hline
\end{tabular}}
\caption{Summary of the best-fit physical parameters including the PWN particle properties for the evolutionary system considering PWN+PSR contributions to the Fermi--LAT emission.}\label{tab:inputoutputgelfand}
\end{table*}

\section{Conclusion}
In order to determine the potential for PWNe to efficiently accelerate particles, we must be able to constrain the synchrotron cut-off and MeV-GeV $\gamma$-ray shapes accurately. In particular, the particle cut-off energy and maximum energy as well as the properties of the ambient photon fields determine these shapes. For B0453-685, the radiative models make use of varying values for these parameters, which are listed for convenient comparison in Table~\ref{tab:naima_ev}. Despite the variations in the model techniques and predicted properties, both models predict a synchrotron cut-off from B0453-685 just beyond the {\it Chandra} X-ray energy range (e.g., Figures~\ref{fig:mybroadbandpwnmodel} and \ref{fig:compare-gelfand-seds}). However, the Chandra X-ray spectrum for the PWN is very hard and does not indicate a cut-off. If we want to obtain adequate constraints on the synchrotron cut-off and to understand the potential for a pulsar contribution that is predicted to peak in the MeV band, we need observations in the X-ray range beyond {\it Chandra} or in the sub-50\,MeV $\gamma$-ray band. 

\begin{table}[tbh!]
\centering
\begin{tabular}{| c || c c |}
\hline
\ Property & NAIMA Model & Evolutionary Model \\
\hline 
\hline 
\  Photon fields & CMB & CMB, X-ray  \\
\hline 
\ $E_b$ (GeV) & 350 & 70 \\
\hline 
\ $E_{max}$ (PeV) & 0.3 (fixed) & 0.73 \\
\ Index1 & 0.88 & 1.3 \\
\ Index2 & 2.05 & 2.4 \\
\hline 
\end{tabular}
\caption{The major differences in the competing broadband models, NAIMA (time-independent) and evolutionary (time-dependent). Constraints in the hard X-ray and/or sub-50\,MeV band are needed to characterize the synchrotron cut-off and compare to model predictions.}
\label{tab:naima_ev}
\end{table}


In summary, faint point-like $\gamma$-ray emission is discovered coincident to B0453-685 and a multiwavelength analysis favors an evolved PWN origin where the compression of the PWN has initiated from the return of the SNR reverse shock and leads to distinct particle components in the broadband spectrum. Future hard X-ray and/or MeV observations to further constrain the PWN and pulsar properties are needed. Finally, this system constitutes only the second extragalactic Fermi-detected PWN, after the GeV and TeV detection of N~157B \citep{lmc2016}.

\end{document}